\documentclass[showpacs,twocolumn,aps,floatfix,superscriptaddress]{revtex4-1}
\usepackage{amsmath,amssymb,eucal,graphicx,bm}

\begin{document}
\title{Lorentz Gas at Positive Temperature}
\author{L. D'Alessio}
\author{P. L. Krapivsky}
\affiliation{Department of Physics, Boston University, Boston, MA 02215, USA}
\begin{abstract}
We investigate the evolution of a particle in a Lorentz gas where the background scatters move and collide with each other. 
As in the standard Lorentz gas, we assume that the particle is negligibly light in comparison with scatters. 
We show that the average particle speed grows in time as $t^{\lambda/(4+\lambda)}$ in three dimensions if the particle-scatter potential diverges as $r^{-\lambda}$ in the small separation limit. The particle displacement exhibits a universal growth, linear in time and the average speed of the atoms. Surprisingly, the asymptotic growth is independent on the gas density
and the particle-atom interaction. The velocity and position distributions approach universal scaling
forms which are non-Gaussian. We determine the velocity distribution in arbitrary dimension and
for arbitrary interaction exponent $\lambda$. 
For the hard-sphere particle-atom interaction, we compute the position distribution and the joint velocity-position distribution.
\end{abstract}
\maketitle

The Boltzmann equation \cite{BOLT} is the basic tool in elucidating the properties of transport phenomena. The non-linear integro-differential Boltzmann equation is so formidable, however, that apart from the equilibrium Maxwell-Boltzmann distribution \cite{MAX} there are essentially no solutions to the Boltzmann equation \cite{TM80}. The standard Lorentz gas model where a point particle is elastically scattered by immobile hard spheres is described by the Lorentz-Boltzmann equation \cite{LG} which is linear and, not surprisingly, amenable to analytical treatments. The Lorentz gas has played an outstanding role in concrete calculations (e.g. of the diffusion coefficient) and in the conceptual development of kinetic theory \cite{LinearLG,fluid}. Yet the very applicability of the Boltzmann framework to the Lorentz gas is questionable --- when the scatters are fixed, the molecular chaos assumption underlying the Boltzmann equation is hard to justify \cite{LinearLG,fluid,HardBallGas,book}. 

If, however, the background particles (atoms for short) move and collide with each other, the molecular chaos assumption holds in the dilute limit and the (properly generalized) Lorentz-Boltzmann equation must be applicable as long as the mass of the point particle is infinitesimally small so that it does not affect the motion of atoms. Surprisingly this model has not been studied in the context of kinetic theory until very recently \cite{italy} and many aspect of it are still unclear (e.g. density profile). This problem is also reminiscent of the model proposed by Fermi \cite{F} to explain the acceleration of interstellar particles which has been mostly studied using methods of dynamical systems (see e.g. \cite{LL} and references therein). 

The emerging behavior of the particle in our model is drastically different from the case of the standard Lorentz gas with immobile atoms. 
The average particle velocity and and displacement grow linearly in time. 
Further, after a re-scaling with respect to the average quantities, the velocity and displacement distributions approach universal forms which are not Gaussian.

The ratio of masses determines the particle equilibrium velocity. When this ratio is zero the equilibrium particle's velocity is infinite and on the quest to equilibration the particle speed increase indefinitely in time. 
Note that in this case the particle carries no kinetic energy and no momentum so the conservation of energy and momentum do not constrain the particle's velocity.
A light particle will eventually thermalize with the background atoms at some finite velocity and the velocity and displacement distribution will become Gaussian. Taking the limit of a negligible light particle allow us to push the equilibration time to infinity and to observe interesting not equilibrium behaviors. Our calculation correctly reproduces the behavior of systems with a mass ratio not strictly zero up to some characteristic time where the onset of equilibration appears \cite{LP}. 

Let us first analyze the velocity distribution. Suppose the atoms are hard spheres of radius $a$. The particle velocity distribution $f({\bf v},t)$ satisfies the Lorentz-Boltzmann equation
\begin{equation}
\label{B-d}
\frac{\partial f}{\partial t}=\int d{\bf u}\,P({\bf u})\,ga^{d-1}\!
\int\! \mathcal{D}{\bf e}\,[f({\bf v}',t)-f({\bf v},t)]\,.
\end{equation}
Here $P({\bf u})=\rho(2\pi T)^{-d/2}e^{-u^2/2T}$ is the Maxwell-Boltzmann velocity distribution of the background gas ($d$ is the spatial dimension, $\rho$ the gas density, $T$ the temperature, and the atomic mass was set to unity), ${\bf e}$ is the unit vector pointing to the position of the particle at the moment when it hits the sphere and $\mathcal{D}{\bf e}$ is the integration measure over angular coordinates.  
The measure $\mathcal{D}{\bf e}$ additionally depends on the relative velocity ${\bf g}={\bf u}-{\bf v}$; for the hard-sphere gas $\mathcal{D}{\bf e} = \frac{({\bf g}\cdot{\bf e})}{g}\,\theta({\bf g}\cdot{\bf e})\,d{\bf e}$, where $\theta(\cdot)$ is the Heaviside step function and $d{\bf e}$ is the standard angular integration measure. The post-collision velocity ${\bf v}'$  of the particle can be expressed via ${\bf v}, {\bf e}$, and ${\bf g}$:
\begin{equation}
\label{vv}
{\bf v}' = {\bf v} + 2{\bf e}({\bf g}\cdot{\bf e})\,.
\end{equation}
Equation \eqref{B-d} is applicable in the diluted limit when the volume fraction occupied by atoms is small: $\rho\,a^d\ll 1$. 

The collision integral in Eq.~\eqref{B-d} can be simplified in the long time limit assuming that the velocity distribution is isotropic and that the particle velocity greatly exceeds the the typical velocity of background atoms, $v \gg \sqrt{T}$.  
The last assumption will allow us to expand the collision integral in a power series of the small parameter $\sqrt{T}/v$. 
It is worth noting that an initial velocity $v\ll\sqrt{T}$ will become of order $\sqrt{T}$ after a single collision (see Eq.~\eqref{vv}).
 
First we treat $f({\bf v})$ as a function of $V={\bf v}\cdot{\bf v}$.
Squaring \eqref{vv} we get $V' = V+4({\bf u}\cdot{\bf e})({\bf g}\cdot{\bf e})$. Using this result and expanding $f({\bf v}')=f(V')$ into a Taylor series up to the second order we obtain
\begin{equation*}
f(V')=f(V)+4({\bf u}\cdot{\bf e})({\bf g}\cdot{\bf e})\,\frac{\partial f}{\partial V}
+ 8({\bf u}\cdot{\bf e})^2({\bf g}\cdot{\bf e})^2\,\frac{\partial^2 f}{\partial V^2}\,.
\end{equation*}

We now insert this expression back into Eq.~\eqref{B-d} and we are left with integrals over $\mathcal{D}{\bf e}$ and $d{\bf u}$. 
Using symmetry we deduce \cite{LP} the dependence of the angular integrals on $\mathbf{u}$ and $\mathbf{g}$:
\begin{equation*}
\begin{split}
\int\mathcal{D}{\bf e}\,(\mathbf{u}\cdot\mathbf{e})(\mathbf{g}\cdot\mathbf{e}) & =A(\mathbf{u}\cdot\mathbf{g})\\
\int\mathcal{D}{\bf e}\,(\mathbf{u}\cdot\mathbf{e})^{2}(\mathbf{g}\cdot\mathbf{e})^{2} & =\frac{dB-A}{d-1}\,(\mathbf{u}\cdot\mathbf{g})^{2}+\frac{A-B}{d-1}\, g^{2}u^{2}
\end{split}
\end{equation*}
where $A$, $B$ are constants defined by integrals: 
\begin{equation*}
A=\frac{1}{g^{2}}\int\mathcal{D}{\bf e}\,(\mathbf{g}\cdot\mathbf{e})^{2}\,,\quad
B=\frac{1}{g^{4}}\int\mathcal{D}{\bf e}\,(\mathbf{g}\cdot\mathbf{e})^{4}\,.
\end{equation*}
The integrals over ${\bf u}$ can be computed by taking into account that the velocity of atoms $u\sim\sqrt{T}$ is asymptotically negligible with respect to the particle velocity. We thus expand the integrand into Taylor series of $u/v$. 
Completing the integration reduces \cite{LP} the integro-differential Lorentz-Boltzmann equation to the partial differential equation 
\begin{equation} 
\frac{\partial f}{\partial\tau}=d\,\frac{\partial f}{\partial v}+v\,\frac{\partial^{2}f}{\partial v^{2}}\,,\,\,\tau=2a^{d-1}A\rho Tt.
\label{final_v}
\end{equation}
Interestingly, the constant $B$ drops from the final equation; the constant $A$ is essentially irrelevant as it is absorbed into the new time variable $\tau$. Repeating this procedure it is possible to show \cite{LP} that higher terms in the Taylor expansion of $f(V')$ leads to terms in Eq.~\eqref{final_v} that are suppressed by the powers of the ratio $\sqrt{T}/v$ and are thus negligible in the long-time limit.       
Equation \eqref{final_v} is a much simpler equation than Eq.~\eqref{B-d} and can be solved exactly via Laplace transform \cite{LP}.
In particular it is possible to show that for a delta-function initial condition, $f(v,\tau=0)=\delta(v-v_0) / (\Omega_d \, v_0^{d-1})$
where $\Omega_d=\frac{2\pi^{d/2}}{\Gamma(d/2)}$ is the area of the unit sphere in $d$ dimension, the asymptotic solution is \cite{LP}
\begin{equation}
f(v,\tau)=\frac{1}{\Omega_d \Gamma(d)} \, \frac{e^{-v/\tau}}{\tau^d} 
\label{asymptotic_solution}
\end{equation}
The same asymptotic behavior \eqref{asymptotic_solution} holds for any initial condition 
that decays to zero exponentially or faster, otherwise the long time asymptotic behavior is still given
by Eq.~\eqref{asymptotic_solution} apart from the tail region which is determined by the initial condition \cite{LP}.

In two dimensions, Eqs.~\eqref{final_v}--\eqref{asymptotic_solution}
have been derived in Ref.~\cite{italy} in the realm of a stochastic model for Fermi's acceleration. Even earlier, the exponential velocity distribution was found to occur in another stochastic model for Fermi's acceleration \cite{poland} in which a particle is bouncing in a deforming irregular container of fixed volume; the velocity distribution becomes exponential independently of the container's
shape and deformation protocol \cite{box}. 

We have also considered the case where the interaction between the particle and an atom separated by distance $r$ can be described by a potential function $U(r)$ that diverges as $U\simeq  \epsilon (r/r_0)^{-\lambda}$ in the small separation limit.  In this situation the Lorentz-Boltzmann equation reduces to a kinetic equation (analog to Eq.~\eqref{final_v}) depending only on the interaction exponent $\lambda$ \cite{LP}: 
\begin{equation}
\frac{\partial f}{\partial\tau}=v^{-\gamma}\left[\left(d-\gamma\right)\frac{\partial f}{\partial v}+v\,\frac{\partial^{2}f}{\partial v^{2}}\right],
\label{kinetic}
\end{equation}
where $\tau=2A\left(r_{0}\epsilon^{1/\lambda}\right)^{d-1}\rho Tt$ and $\gamma=2(d-1)/\lambda$.

Equation \eqref{kinetic} admits a scaling solution of the form
\begin{equation}
f=\tau^{-\Lambda d}\Phi(w),\quad w=v\tau^{-\Lambda}\,,\quad\Lambda\equiv(1+\gamma)^{-1}\,.
\label{scaling-3d-form}
\end{equation}
Plugging \eqref{scaling-3d-form} into \eqref{kinetic} we obtain
an ordinary differential equation for $\Phi(w)$ which is solved to yield 
\begin{equation}
\Phi(w)=C\,\exp\!\left\{ -\Lambda^{2}w^{1/\Lambda}\right\}\,,\quad
C = \frac{\Lambda^{2\Lambda d - 1}}{\Omega_d\,\Gamma(\Lambda d)}\,.
\label{scaling-3d-lambda}
\end{equation}
Thus the asymptotic growth, $\langle v\rangle \sim \tau^\Lambda$,  of the average speed and the scaled velocity distribution have universal behaviors that are solely determined by the interaction exponent $\lambda$ and the spatial dimensionality $d$. 

We now investigate the spatial behavior of the particle which are the main results of this paper. We first use a heuristic argument. 
In the case of the hard-sphere interaction, the average speed is $v\sim \tau\sim \rho a^{d-1} Tt$, the  mean-free path is $\ell\sim (\rho a^{d-1})^{-1}$, a time interval between collisions is $\Delta t\sim \ell/v$, and hence the total number of collisions during the time interval $(0,t)$ is 
\begin{equation*}
\mathcal{N}\sim \frac{t}{\Delta t}\sim \frac{v t}{\ell} \sim \frac{T t^2}{\ell^2}\,.
\end{equation*}
Using the standard random walk arguments we estimate the particle typical displacement
\begin{equation}
\label{displ}
r_{\rm typ}\sim \ell\sqrt{\mathcal{N}}\sim \sqrt{T}\, t\,.
\end{equation}
The striking feature is that the displacement is asymptotically independent on the density of atoms and their size. This heuristic arguments can be extended to the case when the particle-atoms interaction is described by a potential and one finds  \cite{LP}  that the displacement obeys the same growth law \eqref{displ}.

To achieve a quantitative understanding of the spread of the particle we must analyze the joint distribution function $f(\mathbf{r},\mathbf{v},t)$ since the spatial distribution function $N(\mathbf{r},t)$ alone does not obey a closed equation. 

For concreteness, consider the one-dimensional case. This setting is physically dubious as the particle is caged between two adjacent atoms, so the molecular chaos assumption (that is, the lack of correlations between pre-collision velocities) underlying the Boltzmann approach is certainly invalid in one dimension. Nevertheless, a Lorentz-Boltzmann equation can be written and it actually describes the situation when in each collision the scattering occurs with a certain probability (otherwise the particle and an atom pass through each other) \cite{book,R}. Then the joint distribution evolves according to
\begin{equation}
\frac{\partial f}{\partial t}+v\,\frac{\partial f}{\partial x}=2 \rho T \left(\frac{\partial f}{\partial v}+v\,\frac{\partial^{2}f}{\partial v^{2}}\right).
\label{kinetic-1d}
\end{equation}
The right-hand side of Eq.~\eqref{kinetic-1d} is the same as in Eq.~\eqref{final_v} (after using $d=1$ and the proper value $A=1$ in one dimension), so the right-hand side is asymptotically exact when $v\gg\langle u\rangle\sim\sqrt{T}$. Further, in this $v\gg \sqrt{T}$ limit the convective term ($v \nabla f$) can be replaced by the diffusion term ($-D\nabla^{2}f$). This is evident if we recall that in the standard Lorentz gas, the particle undergoes diffusion in the \textit{hydrodynamic limit} (see e.g. \cite{LinearLG,fluid,HardBallGas,book}). In our model, in the large time limit the particle experiences many collisions in a short time interval when the speed remains almost constant. The separation between the time scale at which diffusion appears (few collisions) and the time scale at which the particle speed changes appreciably allows us to replace the convective term by the diffusion term.  
In one dimension $D=\frac{v}{2 \rho}$ \cite{book} and thus the governing equation becomes
\begin{equation}
\label{hydrodynamic-1d}
\frac{\partial f}{\partial \tau}
=\frac{\partial f}{\partial v}+v\,\frac{\partial^2 f}{\partial v^2}
+ \frac{|v|}{4\rho^2 T}\,\frac{\partial^2 f}{\partial x^2}
\end{equation}
where we use again $\tau=2\rho T t$ as the time variable. Note that the diffusion coefficient is proportional to velocity and thus on average it linearly grows in time (see Eq.~\eqref{asymptotic_solution}). Equation \eqref{hydrodynamic-1d} is asymptotically exact in the hydrodynamic limit, $v\gg \sqrt{T}$; the full integro-differential Boltzmann equation is required for the description of the early stage of the time evolution when the particle has experienced only few collisions.

In the long time limit, the joint distribution function $f(x,v,t)$ should approach the scaling form
\begin{equation}
\label{scaling-f}
f(x, v, t) \simeq \frac{1}{4 x_{*} v_{*}}\, F(X, V),\quad X=\frac{x}{x_*},\quad V=\frac{v}{v_*}
\end{equation}
where $x_*=\sqrt{T}t$ and $v_*=\tau$.
The velocity and displacement reflection symmetry \cite{LP} allows us to limit ourself to the quadrant $V>0, X>0$.
The normalization condition for $f(x,v,t)$ is recast in $\int_0^\infty dX \int_0^\infty dV\,F(X, V)=1$ 
which explains the factor $1/4$ in the scaling ansatz \eqref{scaling-f}.

It is convenient to study Eq.\eqref{hydrodynamic-1d} on the entire $x$-axis with $v\geq 0$. Performing the Laplace transform in $v$ and the Fourier transform in $x$ we find that the transformed joint distribution
\begin{equation}
\label{LF_def}
g(k, q, \tau)= \int_{-\infty}^\infty dx\,e^{iqx}\int_0^\infty dv\,e^{-vk}\,f(x,v,\tau)
\end{equation}
satisfies
\begin{equation}
\label{LF_kinetic}
\frac{\partial g}{\partial\tau} + \left(k^2-Q^2\right)\frac{\partial g}{\partial k} = -k\,g,
\quad Q^2\equiv \frac{q^2}{4\rho^2 T}
\end{equation}

\begin{figure}
\begin{tabular}{cc}
\includegraphics[width=0.9\columnwidth]{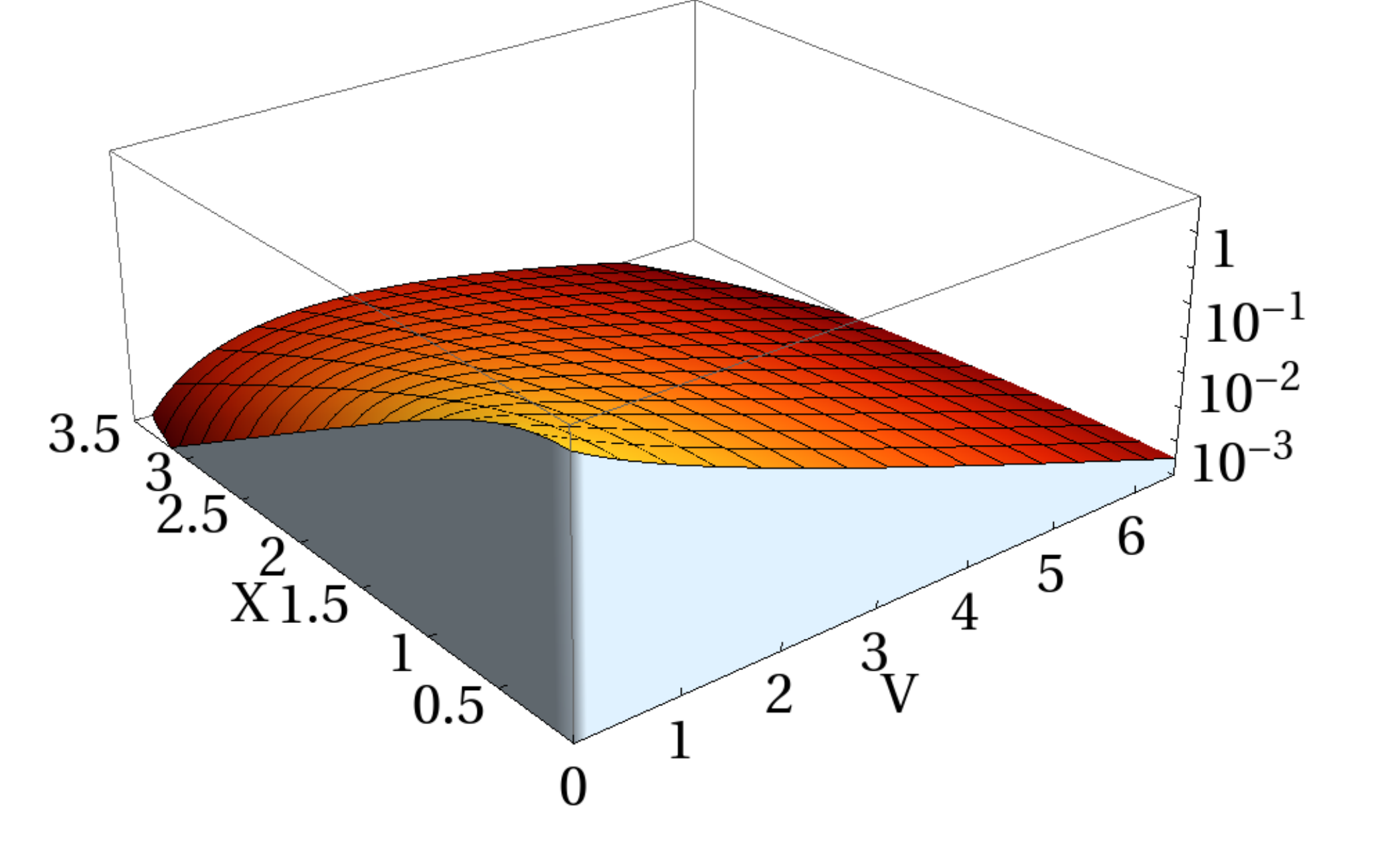} \\
\includegraphics[width=0.9\columnwidth]{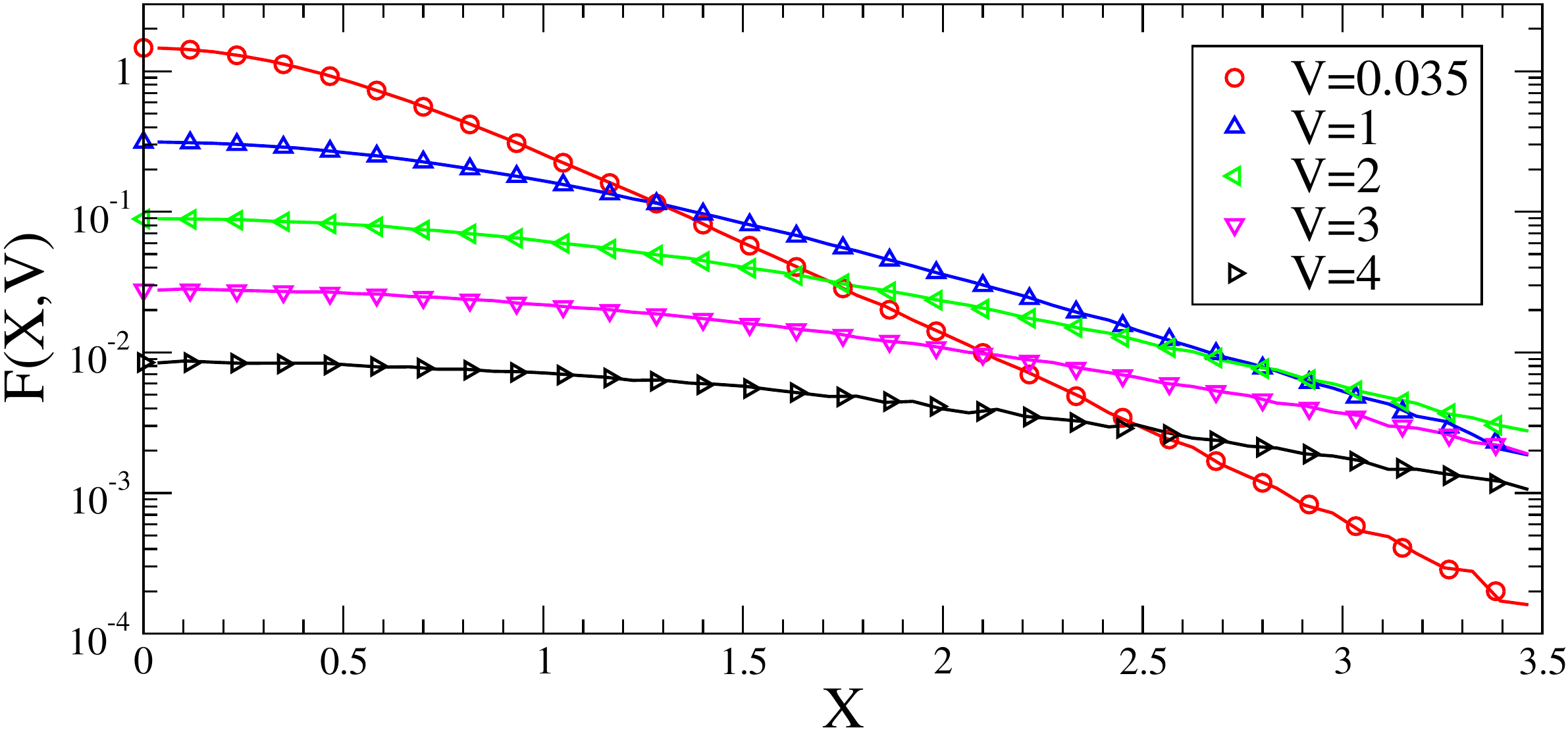}
\end{tabular}
\caption{$F(X,V)$ for hard spheres gas in $1d$, Eq.~\eqref{joint_1d}. (Top panel) 3D-plot.  
(Bottom panel) Values along the lines of fixed $V=0.0035,1,2,3,4$. The continuous lines are obtained from the numerical simulations 
while the symbols represent the values obtained by computing the integral \eqref{joint_1d}.}
\label{theory}
\end{figure}

The exact solution of this linear hyperbolic partial differential equation can be found using the method of characteristics to yield \cite{LP} 
\begin{equation*}
g(k,Q,\tau)= \frac{1}{\cosh s+\frac{k}{Q}\,\sinh s}\,\,
g_0\!\left(\frac{k+Q\tanh s}{1+\frac{k}{Q}\,\tanh s}, Q\right)
\end{equation*}
where $s=Q\tau$ and $g_0(k,Q)=g(k, Q, \tau=0)$. Taking the long-time limit (which, for rapidly decaying initial conditions, is equivalent to choose $g_0$=1 \cite{LP}) and explicitly computing the inverse Laplace transform we obtain scaling function $F(X,V)$ 
\begin{equation}
\label{joint_1d}
F(X,V) = \int_{-\infty}^\infty \frac{ds}{\pi} \, e^{-isX} \frac{s e^{-Vs\coth s}}{\sinh s}
\end{equation}
We could not compute the integral \eqref{joint_1d} in a closed form, so we determined it numerically (Fig.~\ref{theory}). The spatial distribution ($N(X)=\int_0^\infty dV F(X,V)$) admits an explicit expression
\begin{equation}
\label{NX}
N(X) = [\cosh R_1]^{-1}\,,\quad R_1=\frac{\pi}{2}\, X
\end{equation}

\begin{figure}
\includegraphics[width=0.9\columnwidth]{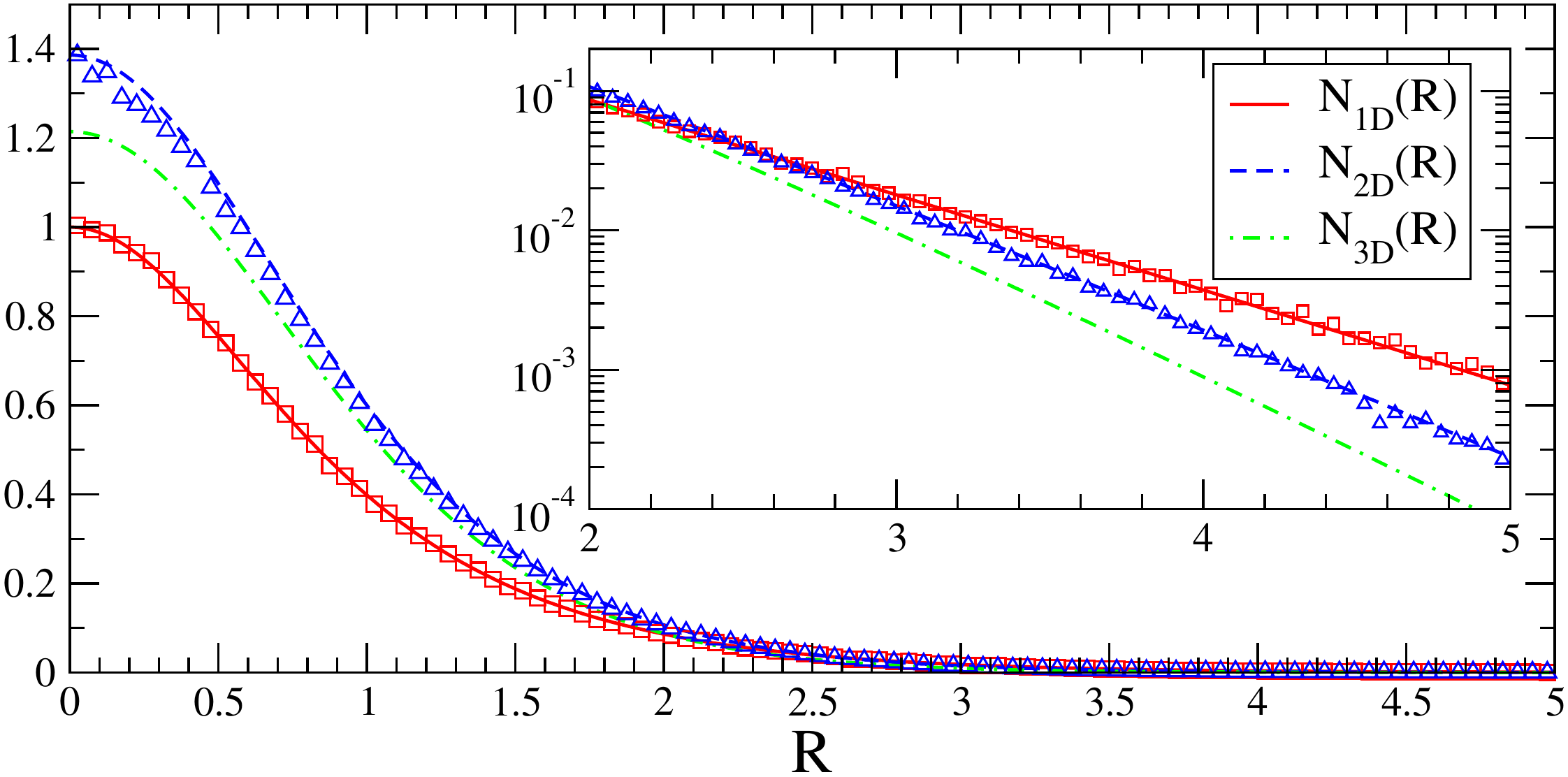}
\caption{Density profile for the hard sphere gas vs. the rescaled variable $R=r/\sqrt{T} t$.
The numerical simulations (symbols) in $d=1,2$ are compared with the theoretical predictions (continuous lines).
The theoretical prediction for $d=3$ is also shown.}
\label{r}
\end{figure}

The same approach holds in any dimension for a hard-sphere particle-atom interaction \cite{LP}. For example, in three dimensions 
\begin{eqnarray*}
&&F(R,V) =  \frac{3}{\pi R} \int_0^\infty ds\,s^4\,\frac{\sin(\sqrt{3}sR)}{(\sinh s)^3} \,e^{-Vs\coth s} \\
&& N(R)   = \frac{3\sqrt{3}}{2}\, \frac{\big(R_3+\frac{\pi^2}{4R_3}\big)\tanh R_3-2}{\cosh R_3}\,,
~ R_3=\frac{\pi\sqrt{3}}{2}\, R
\end{eqnarray*}

For the hard sphere gas in one and two dimensions, the inhomogeneous Lorentz-Boltzmann equation \eqref{B-d} was simulated by stochastically updating the velocities and positions of $10^{8}$ and $10^{6}$ particles respectively \cite{LP}. The velocity distribution is in excellent agreement with the exponential scaling form. The density profiles are shown in Fig.~\ref{r}.
Both in one and two dimensions there is excellent agreement with the theoretical prediction on the full range of the spatial coordinate.

In conclusion we have analyzed the behavior of a very light particle in an equilibrium background gas. We have shown that in the long-time limit, the average particle displacement grows linearly with time and proportionally to the thermal velocity of the background atoms --- the density of the gas, the size of atoms, and the details of the interaction between the particle and the atoms do not affect the asymptotic. The average particle velocity also grows in a rather universal way and the scaled velocity distribution approaches a scaling form which is generically non-Gaussian (the only exception is when the particle-atoms interaction is described by a Maxwell potential, $\gamma=1$ in Eq.~\eqref{kinetic}).
For the hard-sphere particle-atom interaction in arbitrary dimensions, we have computed the asymptotically exact velocity distribution, position distribution and joint velocity-position distribution using a combination of Fourier and Laplace transforms. Our analytic solutions explicitly show the lack of factorization: The joint distribution $F(R,V)$ is not a product of the density $N(R)$ and the velocity distribution $\Phi(V)$.

Our theoretical predictions are in perfect agreement with the numerical simulations providing strong evidence that our simulation scheme is correct and that the simplification of the collision integral and the replacement of the convective term by effective diffusion are indeed asymptotically exact in the limit when the particle velocity greatly exceeds the thermal velocity of atoms.

The Lorentz model was originally suggested \cite{LG} as an idealized model of electron transport. Perhaps the most interesting extension of the present work is to analyze the quantum version of our model.

\vskip 0.2cm
We thank A. Polkovnikov for fruitful discussions. We 
acknowledge support from NSF grant CCF-0829541 (PLK) and DOE grant DE-FG02-08ER46512 (LD'A).

\end{document}